\documentclass[8pt]{article}
\usepackage{epsfig}

\DeclareMathVersion{bold}

\begin{document}

\newcommand{\mmatrix}{$\mathbf{A}$ }
\newcommand{\csmatrix}{$\mathbf{C}_s$ }
\newcommand{\bmatrix}{$\mathbf{B}$ }

\title{Separation of 
correlated astrophysical sources using multiple-lag data
covariance matrices}

\author{L. Bedini$^{1}$,
     D.  Herranz$^{1}$, E.  Salerno$^{1}$, \\C.  Baccigalupi$^{2}$, E.E.  
     Kuruo\u glu$^{1}$, A.  Tonazzini$^{1}$ \\
     \\
 $^{1}$Istituto di Scienza e Tecnologie dell'Informazione,
      CNR, \\Area della Ricerca di Pisa,
      via G. Moruzzi 1, 56124 Pisa, Italy. \\
 $^{2}$International School for Advanced Studies,\\
      via Beirut 4, 34014 Trieste, Italy. \\}

\date{}

\maketitle

\begin{abstract}
This paper deals with a source separation strategy based on 
second-order statistics, namely, on data covariance matrices estimated 
at several lags.  In general, ``blind'' approaches to source 
separation do not assume any knowledge on the mixing operator; 
however, any prior information about the possible structure of the 
mixing operator can improve the solution.  
Unlike ICA blind separation approaches, 
where mutual independence between the sources is assumed, our method 
only needs to constrain second-order statistics, and is effective even 
if the original sources are significantly correlated.  Besides the 
mixing matrix, our strategy is also capable to evaluate the source 
covariance functions at several lags.  Moreover, once the mixing 
parameters have been identified, a simple deconvolution can be used to 
estimate the probability density functions of the source processes.  To 
benchmark our algorithm, we used a database that simulates the one 
expected from the instruments that will operate onboard ESA's Planck 
Surveyor Satellite to measure the CMB anisotropies all over the 
celestial sphere.  The assumption was made that the emission spectra 
of the galactic foregrounds can be parametrised, thus reducing the 
number of unknowns for system identification to the number of the 
foreground radiations.  We performed separation in several sky 
patches, featuring different levels of galactic contamination to the 
CMB, and assuming several noise levels, including the ones derived 
from the Planck specifications.  

\end{abstract}


\section{Introduction}

Separating the individual 
radiations from the measured signals is a common problem in 
astrophysical data analysis \cite{tegmark00}.  As an example, in 
cosmic microwave background anisotropy surveys, the cosmological 
signal is normally combined with foreground radiations from both extragalactic and 
galactic sources, such as the Sunyaev-Zeldovich effects from clusters 
of galaxies, the effect of the individual galaxies, the emission from 
galactic dust, the galactic synchrotron and free-free emissions.  If 
one is only interested in estimating the CMB anisotropies, the 
interfering signals can just be treated as noise, and reduced by 
suitable cancellation procedures.  However, the foregrounds have an 
interest of their own, and it could be useful to extract all of them 
from multichannel data, by exploiting their different emission 
spectra.

Some authors \cite{hobson98}\cite{bouchet99} have tried to 
extract a number of individual radiation data from measurements on 
different frequency channels, assuming that the physical mixture model 
is perfectly known.  Unfortunately, such an assumption is rather 
unrealistic and could overconstrain the problem, thus leading to 
unphysical solutions.  Attempts have been made to avoid this 
shortcoming by introducing criteria to evaluate {\em a posteriori} the 
closeness to reality of the mixture model and allowing individual 
sources to be split into separate templates to take spatial parameter 
variability into account \cite{jones00}\cite{barreiro03}.

A class of techniques capable of estimating the source signals as well 
as identifying the mixture model has recently been proposed in 
astrophysics 
\cite{baccigalupi00}\cite{maino02}\cite{baccigalupi02}\cite{delabrouille02}. 
In digital signal processing, these 
techniques are referred to as blind source separation (BSS) and rely on 
statistical assumptions on the source signals.  In particular, mutual 
independence and nongaussianity of the source processes are often 
required \cite{hyvarinen00}.  This totally blind approach, 
denoted as independent component analysis (ICA), has already given 
promising results, proving to be a valid alternative to assuming a 
known data model.  On the other hand, most ICA algorithms do not 
permit to introduce prior information.  Since all available 
information should always be used, semi-blind techniques are being 
studied to make astrophysical source separation more flexible with 
respect to the specific knowledge often available in this type of 
problem \cite{kuruoglu03}.  Moreover, the independence 
assumption is not always justified; if there is evidence of 
correlation between pairs of sources, it should be made possible to 
take this information into account, thus abandoning the strict ICA 
approach.

The first blind technique proposed to solve the separation problem in 
astrophysics \cite{baccigalupi00} was based on ICA, and allowed 
simultaneous model identification and signal estimation to be 
performed.  The independence requirement was fulfilled by taking the 
statistics of all orders into account, as in all ICA methods presented 
in the literature (see for example 
\cite{comon94}\cite{cardoso98}\cite{hyvarinen00}).

The problem of estimating all the model parameters and source signals 
cannot be solved by just using second-order statistics, since these 
are only able to enforce uncorrelation.  However, this has been done 
in special cases, where additional hypotheses on the spatial 
correlations or, equivalently, on the spectra of the individual 
signals are assumed 
\cite{tong91}\cite{belouchrani97}\cite{delabrouille02}.  
As will be clear in the following, within the framework of any noisy 
linear mixture model, the data covariance matrix at a particular lag 
is related to the source covariance matrix at the same lag, the mixing 
matrix, and the noise covariance matrix.  If there is a sufficient 
number of lags for which the source covariance matrices are not null, 
then it is possible to identify the model parameters by estimating the 
data covariance matrices from the observed data.  Indeed, assuming to 
know the noise covariance matrix, we are able to write a number of 
relationships from which the unknown parameters can be estimated.  
This is what is done by the second-order blind identification (SOBI) 
algorithm presented in \cite{belouchrani97}.  SOBI, however, 
relies on joint digonalization of covariance matrices at different 
lags, which is only applicable in the case of uncorrelated source 
signals. In our approach, we assumed that the mixing matrix can be 
parametrised. This allows us to relax the independence assumption, and 
to pursue identification by optimization of a suitable 
function. A further advantage of this 
strategy is that the relevant correlation coefficients between pairs 
of sources can also be estimated.
In the particular case of separating astrophysical foregrounds from 
cosmic microwave background, moreover, the relevant constraints are 
such that the total number of parameters to be estimated can 
substantially be reduced.  This permits to improve the performance of 
our technique.  We will show that, even assuming full covariance 
matrices at different lags, a very fast model learning algorithm can 
be devised, matching the theoretical covariance matrices to the ones 
estimated from the observed data.

The paper is organised as follows.  In Section 
\ref{sec:source_separation}, we formalise the problem and introduce 
the relevant notation.  In Section \ref{sec:parametrization}, we 
describe how the mixing matrix can be parametrised in 
our case.  In Sections \ref{sec:algorithm} and \ref{sec:separation}, 
we describe the methods we used to learn the mixing model and to 
estimate the original sources, respectively.  In Section 
\ref{sec:experiments}, we present some experimental results, with both 
stationary and nonstationary noise. In 
the final section, we give some remarks and future directions.

\section{Problem statement} \label{sec:source_separation}
As usual \cite{hobson98}\cite{baccigalupi00}, 
we assume that each radiation process 
$\tilde{s}_c(\xi,\eta,\nu)$
from the microwave sky has a spatial pattern
$s_c(\xi,\eta)$
that is independent of its frequency spectrum
$F_c(\nu)$:
\begin{equation} \label{eq:factorization}
\tilde{s}_c(\xi,\eta,\nu) = s_c(\xi,\eta)  F_c(\nu)
\end{equation}
\noindent
Here, $\xi$ and $\eta$ are angular coordinates on the celestial sphere, 
and $\nu$ is frequency. The total radiation observed in a certain direction 
at a certain frequency is given by the sum of a number $N$ of 
signals (processes, or components) of the type (\ref{eq:factorization}), 
where subscript $c$ has the meaning of a process index. 
Assuming that the effects of the telescope beam on the angular 
resolution at different measurement channels have been 
equalised (see \cite{salerno00}), the observed signal at $M$ 
different frequencies can be modelled as
\begin{equation} \label{eq:model}
\mathbf{x}(\xi,\eta) = \mathbf{A}\mathbf{s}(\xi,\eta)+\mathbf{n}(\xi,\eta)
\end{equation} 
\noindent
where $\mathbf{x}$=$\{x_d, \  d=1,\ldots , M\}$ 
is the $M$-vector of the observations, 
$d$ being a channel index, $\mathbf{A}$ is an $M\times N$ {\em mixing} matrix, 
\mbox{$\mathbf{s}$ = $\{s_c, \ 
c=1,\ldots ,N\}$} is the $N$-vector of the individual source processes 
and $\mathbf{n}$=$\{n_d, \ d=1,\ldots ,M\}$ is the $M$-vector of 
instrumental noise. The elements of \mmatrix are related to the source spectra 
and to the frequency responses through the following formula:
\begin{equation} \label{eq:a_elements}
a_{dc} = \int F_c(\nu) b_d(\nu) d\nu├╣
\end{equation}
where $b_d(\nu)$ is the instrumental frequency response in the 
\mbox{$d$-th} measurement channel, which is normally known very well.  
If we assume that the source spectra are constant within the passbands 
of the different channels, equation (\ref{eq:a_elements}) can be 
rewritten as
\begin{equation} \label{eq:a_elements2}
a_{dc} = F_c(\nu_d) \int  b_d(\nu) d\nu
\end{equation}
The element $a_{dc}$ is thus proportional to the spectrum of the 
\mbox{$c$-th} source at the center-frequency $\nu_d$ of the 
\mbox{$d$-th} channel.  The separation problem consists in estimating 
the source vector $\mathbf{s}$ from the observed vector $\mathbf{x}$.  
Several estimation algorithms have been derived assuming a perfect 
knowledge of the mixing matrix.  As already said, however, this matrix 
is related to both the instrumental frequency responses, which are 
known, and the emission spectra $F_c(\nu)$, which are normally 
unknown.  For this reason, relying on an assumed mutual independence 
of the source processes $s_c(\xi,\eta)$, some \emph{blind} separation 
algorithms have been proposed 
\cite{baccigalupi00}\cite{maino02}\cite{patanchon03}, which are able 
to estimate both the mixing matrix and the source vector.  Assuming 
that the source signals are mutually independent, the $MN$ mixing 
coefficients can be estimated by finding a linear mixture that, when 
applied to the data vector, nullifies the cross-cumulants of all 
orders.  If, however, some prior information allows us to reduce the 
number of unknowns, the identification problem can be solved by only 
using second-order statistics.  This is the case with our approach, 
which is based on a parametrisation of matrix \mmatrix.  This 
approach, described in Section \ref{sec:algorithm}, does not need a 
strict mutual independence assumption.  Logically, any blind 
separation algorithm is divided into two phases: using the notation 
introduced here, the estimation of \mmatrix will be referred to as 
\emph{system identification} (or \emph{model learning}), and the 
estimation of $\mathbf{s}$ will be referred to as \emph{source 
separation}.  In this paper, we first address aspects related to 
learning, and then give some details on source separation strategies 
derived from standard reconstruction procedures.  Before describing 
our algorithm in detail, we recall here some applicability issues.

\noindent {\bf Source and noise processes.} To estimate the covariance 
matrices from the available data, the source and the noise processes 
must necessarily be assumed stationary.  While CMB satisfies this 
assumption, the foregrounds are not stationary all over the celestial 
sphere.  This assumption can be made for small sky 
patches.  However, depending on the particular sky scanning strategy, noise is 
normally nonstationary, even within small patches, and can also 
be auto-correlated.  The noise covariance function should be known for 
any shift and for any angular coordinate in the celestial sphere.  
Provided that the noise nonstationarity and 
cross-correlation between sources can be neglected, various methods are 
available, both in space and frequency domain, to estimate samples of 
the noise covariance function or, equivalently, of noise spectrum 
\cite{delabrouille02}.  Tackling the space-variant nature of the 
noise process is difficult, and no simple method has been proposed so 
far to this purpose.  In \cite{kuruoglu03} the noise 
variance at each pixel is assumed to be known and a method is 
proposed to estimate 
the mixing matrix and the probability density function of each 
component.  In the present approach, we found experimentally that, if 
a noise covariance map is known, even nonstationary noise can be 
treated.

\noindent {\bf Frequency dependent telescope beams.} The model assumed 
in (\ref{eq:model}) is valid if the telescope radiation patterns are the 
same in all the frequency channels.  As the beams are frequency 
dependent, a way to tackle the problem is to preprocess the observed 
data in order to equalise the resolution on all the measurement 
channels, as in \cite{salerno00}.  This also changes the autocorrelation 
function of each noise process, but in a way that can be exactly 
evaluated.  A different way to tackle 
the problem has been to approach it in the frequency domain 
\cite{hobson98}\cite{delabrouille02}.  Also in these cases, the 
validity of the solution relies on a number of simplifiying 
assumptions, such as the perfect circular symmetry of the telescope 
beams.  Moreover, the actual capability of extrapolating the spectrum at 
spatial frequencies where reduced information is available has still 
to be assessed, especially in the cases where the signal-to-noise 
ratio is particularly low.

\noindent {\bf Structure of the source covariance matrices.} In the 
Planck experiment, the sources of interest are the CMB signal and the 
foregrounds.  While no correlation is expected between the CMB signal 
and foregrounds, some statistical dependence between pairs of 
foregrounds has to be taken into account.  The off-diagonal entries of 
the source covariance matrices related to pairs of correlated sources 
will thus be nonzero, whereas all the remaining off-diagonal elements 
will be zero.  When it is known that some of the cross-covariances are 
close to zero, these can be kept fixed at zero, thus further 
reducing the total number of unknowns.  For instance, 
in a $3 \times 3$ case, if we assume the following structure for the 
source covariance matrix at zero-shift: 
\begin{equation} \label{eq:covarianza}
\mathbf{C}_s(\mathrm{0,0}) = \left(
\begin{array}{c c c }
\sigma_{11}е & 0 & 0 \\
0 & \sigma_{22}е & \sigma_{23}е \\
0 & \sigma_{32}е & \sigma_{33}е 
\end{array}
\right)  ,
\end{equation}
this means that we assume zero or negligible correlations between 
sources 1 and 2 and sources 1 and 3, and the remaining 
cross-covariance $\sigma_{23}е=\sigma_{32}е$ between sources 2 and 3 
is an unknown of the problem, along with the autocovariances 
$\sigma_{ii}е$.  
Note that, for the typical scaling ambiguity of the blind 
identification problem, the absolute values of both the diagonal and 
off-diagonal elements of matrices $\mathbf{C}_s(\tau,\psi)$ have no 
physical significance, while, by calculating ratios of the type
\begin{equation} \label{eq:coeffcorr}
\frac{(\sigma_{ij}е)^{2}е}{\sigma_{ii}е\sigma_{jj}ее}е  ,
\end{equation}
we can actually estimate the correlation 
coefficients between different sources, whatever the values of the 
individual covariances.

\section{Parametrisation of the mixing matrix}
\label{sec:parametrization}
While in a general source separation problem 
the elements $a_{dc}$ are totally unknown, in our case we have some 
knowledge about them.  In fact, the integral in (\ref{eq:a_elements2}) 
is related to known instrumental features and to the emission spectra 
of the single source processes, on which we do have some knowledge.  
As an example, if the observations are made in the microwave and 
millimeter-wave range, the dominant radiations are the cosmic 
microwave background, the galactic dust, the free-free emission and 
the synchrotron (see \cite{dezotti99}).  Another significant 
signal comes from the extragalactic radio sources.  Here we assume 
that the latter has been removed from the data by one of the specific 
techniques proposed in the literature \cite{tenorio99}\cite{cayon00} 
\cite{vielva01}.  As a matter of fact, these 
techniques cannot remove totally the extragalactic point sources, but 
they remove the brightest ones (which are the most important, since 
they significantly affect the the study of the CMB, see 
\cite{vielva01}).  As far as the other signals are concerned, the 
emission spectrum of the cosmic microwave background is perfectly 
known, being a blackbody radiation.  In terms of antenna temperature, 
it is:
\begin{equation}  \label{eq:blackbody}
F_{cmb}(\nu) = \frac{\tilde{\nu}^2 \exp(\tilde{\nu})}
{\left[ \exp(\tilde{\nu})-1 \right]^2 }
\end{equation}
where $\tilde{\nu}$ is the frequency in GHz divided by $56.8$. 
From (\ref{eq:a_elements2}) and (\ref{eq:blackbody}), the column of 
\mmatrix related to the CMB radiation is thus known up to an 
unessential scale factor. For the synchrotron radiation, we have
\begin{equation}  \label{eq:sync}
F_{syn}(\nu) \propto \nu^{-n_s}
\end{equation}
\noindent
Thus, the column of \mmatrix related to synchrotron only 
depends on a scale factor and the spectral index $n_s$. 
For the thermal galactic dust, we have
\begin{equation} \label{eq:dust}
F_{dust}(\nu) \propto \frac{\bar{\nu}^{m+1}}{\exp(\bar{\nu})-1}
\end{equation}
where $\bar{\nu} = h \nu /kT_{dust}$, $h$ is the Planck constant,
$k$ is the Boltzmann constant and 
$T_{dust}$ is the physical dust temperature. 
If we assume a uniform temperature value, the frequency law (\ref{eq:dust}), 
that is, the column of \mmatrix related to dust emission, 
only depends on a scale factor and the parameter $m$.  

The above properties enable us to describe the mixing matrix by means 
of just a few parameters.  As an example, if we assume to have a 
perfectly known source spectrum (such as the one of CMB) and $N-1$ 
sources with one-parameter spectra, the number of unknowns in the 
identification problem is $N-1$ instead of $NM$.

For the sake of simplicity, although other foregrounds (such as SZ 
and free-free) could be taken into account, in our experiments we only 
considered synchrotron and dust
emissions, which are the most significant in the Planck frequency
range.

\section{A second-order identification algorithm} \label{sec:algorithm}

Let us consider the source and noise signals in (\ref{eq:model}) 
as realisations of two stationary vector random processes.
The covariance matrices of these processes are, respectively,
\begin{eqnarray}
\mathbf{C}_s(\tau,\psi) & = & 
\langle \left[ \mathbf{s}(\xi,\eta) - \mu_s \right]
\left[ \mathbf{s}(\xi+\tau,\eta+\psi)-\mu_s \right]^{\mathrm{T}} \rangle  \ , 
 \label{eq:cov_signal} \\
\mathbf{C}_n(\tau, \psi) & = & \langle \left[ \mathbf{n}(\xi, \eta) - 
\mu_{n}е\right] \left[ \mathbf{n}(\xi+\tau, \eta+\psi) - 
\mu_{n}е 
\right]^{\mathrm{T}} \rangle
\label{eq:cov_noise}
\end{eqnarray}
where $\langle .  \rangle$ denotes expectation under the appropriate 
joint probability, $\mu_s$ and $\mu_n$ are the mean vectors of 
processes $\mathbf{s}$ and $\mathbf{n}$, respectively, and the 
superscript $\mathrm{T}$ means transposition.  As usual, the noise 
process is assumed signal-independent, white and zero-mean, with known 
variances.  Thus, for both $\tau$ and $\psi$ equal to zero, 
$\mathbf{C}_n$ is a known diagonal matrix whose elements are the noise 
variances in all the measurement channels, whereas for any $\tau$ or 
$\psi$ different from zero $\mathbf{C}_n$ is the null $M \times M$ 
matrix. 

As already proved \cite{belouchrani97}\cite{barros01}, covariance 
matrices, i.e.  second-order statistics, permit blind separation to be 
achieved when the sources show a spatial structure, namely, when they 
are spatially correlated.  Thus, the mutual independence requirement 
of ICA can be replaced by an equivalent requirement on the spatial 
structure of the signal, and the identifiability of the system is 
assured.  In other words, finding matrices \mmatrix and \csmatrix 
is generally not possible from covariances at zero shift alone; to 
identify the mixing operator, either higher-order statistics or the 
covariance matrices at several nonzero shift pairs $(\tau,\psi)$ must 
be taken into account.  Of course, this is also a requirement on the 
sources, since if the covariance matrices are null for any pair 
$(\tau,\psi)$, identification is not possible.  This aspect will 
become clearer below.

Let us now see our approach to system identification. By exploiting equation 
(\ref{eq:model}), the covariance of the observed data can be written 
as:
\begin{eqnarray} \label{eq:cov_shift0}
\mathbf{C}_x(\mathrm{\tau},\psi)  = & \langle \left[ \mathbf{x}(\xi,\eta) - 
\mu_x \right]
\left[ \mathbf{x}(\xi+\tau,\eta+\psi)-\mu_x \right]^{\mathrm{T}} \rangle & = 
\nonumber \\ 
  & 
  = \mathbf{AC}_s(\mathrm{\tau},\psi)\mathbf{A}^{\mathrm{T}}+\mathbf{C}_n(\tau,\psi) 
  \ .  &
\end{eqnarray}
Since $\mathbf{C}_x(\mathrm{\tau},\psi)$ can be estimated from 
\begin{equation} \label{eq:estimation_cx}
\hat{\mathbf{C}}_{\mathbf{x}}(\tau,\psi) = \frac{\mathrm{1}}{\mathit{N}_p} \sum_{\xi,\eta} 
\left[ \mathbf{x}(\xi,\eta)-\mu_x \right] \left[ 
\mathbf{x}(\xi+\tau,\eta+\psi)-\mu_x \right]^{\mathrm{T}} ,
\end{equation}
where $N_{p}е$ is the number of pixels.  Equation 
(\ref{eq:cov_shift0}) provides a number of independent nonlinear 
relationships that can be used to estimate both \mmatrix and 
$\mathbf{C}_s$.  Obviously, this possibility does not rely on mutual 
independence between the source signals, as required by the ICA 
approach: the only requirement is having a sufficient number of 
nonzero covariance matrices.  In other words, spatial structure can be 
used in the place of mutual independece as a basis for model learning 
and signal separation.  As assumed in the previous section, in this 
particular application the number of unknowns is reduced by 
parametrising the mixing matrix.  This allows us to solve the 
identification problem from the relationships made available by 
Equation (\ref{eq:cov_shift0}) by only using the zero-shift covariance 
matrix, even if some of the sources are cross-correlated.  We 
investigated this possibility in \cite{bedini03}.  In a general case, 
matrices \mmatrix and \csmatrix$(\tau,\psi)$ can be estimated from
\begin{equation} \label{eq:estimation_acs}
\left(\Gamma,\Sigma(\tau,\psi) \right) =
\mathrm{arg \ min}
\scriptstyle{
\left\| 
\mathbf{A}              \mathrm{(}\Gamma)
\mathbf{C}_s            \mathrm{(}\Sigma(\tau,\psi))
\mathbf{A}^{\mathrm{T}} \mathrm{(}\Gamma)
- \hat{\mathbf{C}}_x(\tau,\psi)-\mathbf{C}_n(\tau,\psi)
\right\| } .
\end{equation}
The minimisation is performed over vectors $\Gamma$ 
and $\Sigma$, and for all available values of the pairs $(\tau,\psi)$, 
where $\Gamma$ is the vector of all the parameters defining \mmatrix 
(possibly consisting in all the matrix elements), and $\Sigma(\tau, 
\psi)$ is the vector containing all the unknown elements of matrices 
\csmatrix for every shift pair.  The matrix norm adopted is the 
Frobenius norm.
Our present strategy to find the minimiser in 
(\ref{eq:estimation_acs}) is to 
alternate a componentwise minimisation in $\Gamma$ with fixed 
\csmatrix, and the evaluation of \csmatrix, whose elements 
for each $(\tau,\psi)$ can be calculated exactly 
once \mmatrix is fixed.  A more accurate minimisation strategy is now 
being studied.

From the above scheme, it is clear that for each independent element 
of the matrices $\mathbf{C}_x(\tau,\psi)$ we have an independent 
equation for the estimation of vector $\Gamma$ and of all the vextors 
$\Sigma(\tau,\psi)$.  Since for $(\tau,\psi)=(0,0)$ matrix 
$\mathbf{C}_x$ is symmetric, for zero shift we have $M(M+1)/2$ 
independent equations.  For any other shift pair, $\mathbf{C}_x$ is a 
general matrix and thus, provided that it is not zero, we have $M^2$ 
additional independent equations.  If $N_s$ is the total number of 
nonzero shift pairs generating nonzero data covariance matrices, we 
thus have a total number of $M(M+1)/2 + N_s \cdot M^2 = M[(2N_s + 
1)M+1]/2$ independent equations.  The number of unknowns is at most 
$NM + N(N+1)/2 + N_s \cdot N^2$, in the case where all the elements of 
\mmatrix are unknown and all the source covariance matrices are full, 
that is, all the sources at any shift are correlated to each other.  
Note that, in this worst case situation, if it is $M=N$, we always 
have $N^2$ more unknowns than equations, independently of $N_s$.  As 
soon as we have $M>N$, there is always a number of nonzero shift pairs 
for which we have more independent equations than unknowns to be 
estimated.  This observation gives an idea of the amount of 
information we have available for our estimation problem.  The number 
of independent equations affects the behaviour of the nonlinear 
optimization landscape in (\ref{eq:estimation_acs}).  Qualitatively, 
we can affirm that the more independent equations we have, the more 
well-posed the optimization problem will be.  In particular, it is 
likely that in absence of any prior information about the structure of 
\mmatrix and $\mathbf{C}_s(\tau,\psi)$ having a number of observed 
channels equal to the number of sources always leads to insufficient 
information, independently of the number of shift pairs chosen.  If, 
instead, the number of the available observations is larger than the 
number of sources, the possibility of estimating the unknowns relies 
on the number of shift pairs for which the data covariance matrices 
are nonzero.  The availability of prior information, as in the 
application considered here, can of course alleviate these 
requirements.  For example, if we have a $4 \times 4$ mixing matrix 
only depending on four parameters and only two sources significantly 
correlated, the unknowns to be determined are $4+5+N_s \cdot 6$, by 
using a maximum of $M(M+1)/2 + N_s \cdot M^2$ equations.  This means 
that in this case, as soon as $M=4$, the number of independent 
equations is larger than the number of unknowns even for $N_s=0$.

\section{Signal separation strategy}\label{sec:separation}
Model learning is only the first step in solving the problem of source 
separation.  Although, in principle, one could simply use multichannel 
inverse filtering to recover the source maps, this approach is not 
feasible in practice, for the presence of noise.  In our treatment, 
the data are assumed to be an ergodic process, in order to be able to 
evaluate its statistics from the available sample.  This entails a 
space invariant noise process.  The estimation of 
the individual source maps should be made on the basis of all the 
products of the learning stage.  In our case, we have estimates of the 
mixing matrix and of the source covariance matrices at several shift 
pairs.  In the hypothesis of stationary noise, we could exploit this 
information to implement a multichannel Wiener filter for source 
reconstruction. If the noise is not stationary, a generalized Kalman 
filter should be used. Our point here is on model learning, and thus 
we do not address the separation issues in detail. We only observe 
that a possible Bayesian separation scheme would make use of the 
source probability densities, and these can be estimated from our 
mixing matrix.  Indeed, let us assume that our learning procedure has given a 
good estimate of $\mathbf{A}$.  Let 
$\mathbf{B}$ be its Moore-Penrose generalised inverse.  In our case we 
have $M\geq N$, thus, as is known,
\begin{equation} \label{eq:bmatrix}
\mathbf{B} = \left( \mathbf{A}^{\mathrm{T}} \mathbf{A} \right)^{-1} 
\mathbf{A}^{\mathrm{T}} .
\end{equation}
\noindent
From (\ref{eq:model}) we have
\begin{equation} \label{eq:bmatrix2}
\mathbf{Bx} \mathrm{=} \mathbf{s} \mathrm{+} \mathbf{Bn}
\end{equation}
Let us denote each of the $N$ rows of \bmatrix
as an $M$-vector $\mathbf{b}_{\mathnormal{i}}$, $i = 1, \ldots , N$, 
and consider the generic element $y_i$ of the $N$-vector $\mathbf{Bx}$,
\begin{equation}  \label{eq:y_element}
y_i := \mathbf{b}^{\mathrm{T}}_{\mathnormal{i}} \cdot \mathbf{x} =
\mathnormal{s}_i +  \mathbf{b}^{\mathrm{T}}_{\mathnormal{i}} 
\cdot \mathbf{n} :=
\mathnormal{s_{\mathnormal{i}} + n_{t_i}}
\end{equation}
The probability density function of 
$y_i$, $p(y_i)$, can be estimated from $\mathbf{b}_{\mathnormal{i}}$ 
and the data record $\mathbf{x}(\xi,\eta)$, while the probability density 
function of $n_{t_i}$, $p(n_{t_i})$, is a Gaussian, whose parameters can be 
easily derived from $\mathbf{C}_n$ and 
$\mathbf{b}_{\mathnormal{i}}$. 
The pdf of $y_i$ is the convolution between $p(s_i)$ and $p(n_{t_i})$:
\begin{equation} \label{eq:convolution}
p(y_i)=p(s_i) \ast p(n_{t_i}) .
\end{equation}
From this relationship, $p(s_i)$ can be estimated by deconvolution. 
As is well known, deconvolution is normally an ill-posed problem and, 
as such, it lacks a stable solution. In our case, we can regularise 
it by enforcing smoothness, positivity, and the normalisation 
condition for pdfs.

Any Bayesian estimation approach should exploit the knowledge of the 
source densities to regularise the solution, but these are normally 
unknown.  In the case examined here, the source distributions can be 
efficiently
estimated as summarized above, and the computational cost of otherwise 
expensive Bayesian algorithms can be reduced.  As an example, 
in \cite{kuruoglu03}, the source 
densities are modelled as mixtures of Gaussians, and the related 
parameters are estimated by an independent factor analysis approach (see 
\cite{moulines97}\cite{attias99}).  The 
method we propose here could well be used to fix the source 
densities, thus reducing the overall 
cost of the identification-separation task.

From Equation (\ref{eq:bmatrix2}), it can be seen that the generalised 
inverse solution is already an estimate of the sources, since it is 
composed of the original source vectors corrupted by amplified noise.  
Thus, a simple source estimation strategy could be first to apply 
Equation (\ref{eq:bmatrix2}) and then suitably filter the result, to 
reduce the influence of noise.  In next section, we show some 
experimental results obtained by pseudoinversion of the estimated 
mixing matrix, followed by Wiener filtering of each individual source. 
This strategy would be strictly valid with stationary noise and high 
signal-to-noise-ratio, however, interesting results have been found 
even with strong nonstationary noise. Multichannel Wiener filtering 
for stationary noise and an extended Kalman filter for the 
nonstationary case are now being developed.

\section{Experimental results}\label{sec:experiments}

\begin{figure}
\begin{center} 
\epsfxsize=120mm
\epsffile{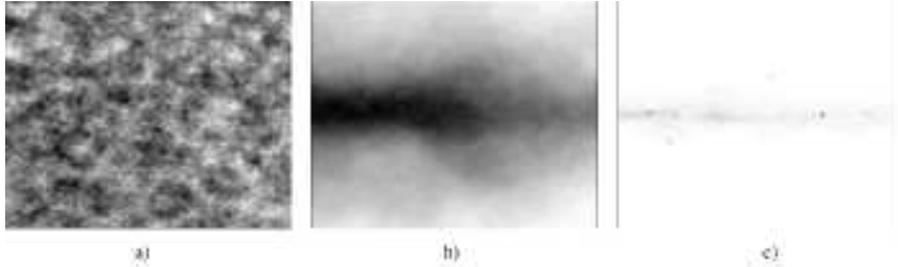}
\end{center}
\caption{Source maps from a $15^{\circ} \times 15^{\circ}$
patch centered at $0^{\circ}$ galactic latitude and 
$40^{\circ}$ galactic longitude, at 100 GHz: 
a) CMB; b) synchrotron; c) thermal dust.}
\label{fig:fig1}
\end{figure}

In this section, we present some results from our extensive 
experimentation with the method described above.  Our data were drawn 
from a data set that somulates the one expected from the Planck 
surveyor satellite (see the Planck 
homepage\footnote{http://astro.estec.esa.nl/SA-general/Projects/Planck/}). 
The source maps we considered were the CMB anisotropy, the galactic 
synchrotron and thermal dust emissions over the four measurement 
channels centred at 30 GHz, 44 GHz, 70 GHz and 100 GHz.  The test data 
maps have been generated by extracting several sky patches at 
different galactic coordinates from the simulated database, scaling 
them exactly according to formulas 
(\ref{eq:blackbody})-(\ref{eq:dust}), generating the mixtures for the 
channels chosen, and adding realisations of Gaussian, signal 
independent, white noise.  Several noise levels have been used, from a 
ten percent to more than one hundred percent of the CMB standard 
deviation.  The range chosen contains noise levels within the Planck 
specifications. Although our method would be only suited for 
uniform noise, we also tried to apply it to data corrupted by 
nonuniform noise, and obtained promising results.

\begin{figure} 
\epsfxsize=84mm
\begin{center}
\epsffile{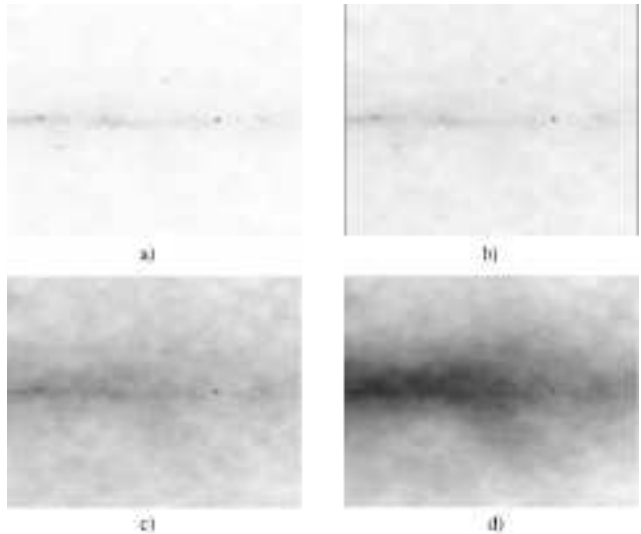}
\end{center}
\caption{
Noisy data maps at a) 100 GHz; b) 70 GHz; c) 44 GHz; d) 30 GHz.}
\label{fig:fig2}
\end{figure}

Within this section, we will divide the results obtained in model 
learning from the results in separation, and the cases with 
stationary noise from those with nonstationary noise. In these latter cases, 
knowledge of a noise variance map is assumed, and the additional problem 
arises of choosing the appropriate noise covariance matrix.

The results from learning are the mixing matrix and the source 
covariance matrices at the shift pairs chosen. From the estimate of 
the mixing matrix, it is also possible to derive the marginal source 
densities, by using relationships (\ref{eq:y_element}) and 
(\ref{eq:convolution}). We have seen that the results under this 
aspect are more sensitive than others to noise, and the choice of the 
regularization parameters is quite critical.

Our separation results are all derived from the application of the 
Moore-Penrose pseudoinverse of the estimated mixing matrix, followed 
by a classical Wiener filtering on each output image. From this 
processing, estimates of the source maps are obtained. Also, 
estimated source power spectra 
can be obtained from either the maps 
or the source autocorrelation matrices. In 
particular, the results we show here are derived from the unfiltered 
pseudoinverse solutions, showing that, although the reconstructed images 
are heavily affected by noise, the derived power spectra can be 
corrected for the theoretical noise spectrum and thus estimated quite 
accurately.

The results presented here will all be related to a single data 
record, derived from a simulated $15^{\circ} \times 15^{\circ}$ 
sky patch centered at $40^{\circ}$ galactic longitude and $0^{\circ}$ 
galactic latitude.  It is to be noted that in such a patch, located on 
the galactic plane, the measured data will be affected by strong 
foreground interference, thus making the problem very difficult to 
solve.  Indeed, many separation approaches experimented so far simply 
fail in proximity of the galactic plane, and they are normally applied 
after masking the all-sky data in the high-interference regions.  It 
is to remark that our method failed with sky patches taken at high galactic 
latitudes, where the only dominant signal is the CMB, and the 
foregrounds are often well below the noise level.  Some other 
techniques, such as ICA (see \cite{baccigalupi00}), did obtain good 
results even in these regions, but the noise levels introduced in 
those cases were much lower than the ones we have used in this work.  
In these regions, furthermore, CMB is almost the only measured 
radiation at the considered frequencies, and is estimated very well 
with all the assigned signal-to-noise ratios.  At lower galactic 
latitudes, conversely, the situation is rather different.  Here, the 
dust emission is stronger than CMB, and separation is strictly 
necessary if CMB is to be distinguished from the foregrounds.  Our 
method performed very well with these data, and all the relevant 
parameters were satisfactorily estimated even with the strongest noise 
components. The noise standard deviation we adopted in the case shown 
here is 30\% the standard deviation of CMB at 100 GHz.  The noise 
level in the other channels has been simply obtained by scaling the 
level at 100 GHz in accordance to the expected Planck sensitivity at 
those frequencies.  For each patch considered, we tried different noise 
levels, up to more than 100\% of the CMB level at 100
GHz, and for each noise level, we performed a Monte Carlo simulation 
with hundreds of different noise realizations. The results of this 
analysis are not shown here, but we can say that no significant bias 
has been found in the results.

\begin{figure} 
\epsfxsize=60mm
\begin{center}
\epsffile{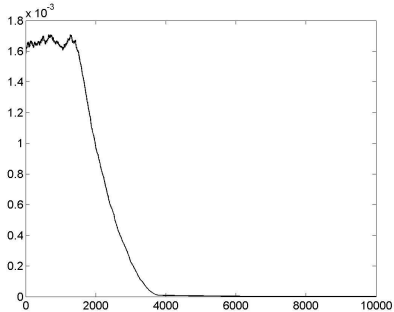}
\end{center}
\caption{Norm of the residual in eq. (\ref{eq:estimation_acs})
as a function of the iteration number.}
\label{fig:fig3}
\end{figure}

In Figure~\ref{fig:fig1}, we show the three source maps we used in the 
situation described above. In this figure and in all the others shown 
here, the grayscale is linear with black corresponding to the maximum 
image value.
We assigned the sources $s_1$ to CMB, $s_2$ to synchrotron and $s_3$ 
to dust, and the signals $x_1$, $x_2$, $x_3$ and $x_4$ to the 
measurement channels at 100, 70, 44 and 30 GHz, respectively.  
Therefore, the first, second, and third columns of the mixing matrix 
will be related to CMB, synchrotron and dust, respectively, and the 
first, second, third, and fourth rows of the mixing matrix will be 
related to the 100 GHz, 70 GHz, 44 GHz, and 30 GHz channels, 
respectively.  The mixing matrix, $\mathbf{A}_o$, has been derived from 
equations (\ref{eq:blackbody})-(\ref{eq:dust}), with spectral indices 
$n_s=2.9$ and $m=1.8$ (see for example \cite{banday91} and 
\cite{giardino02}):
\begin{equation}  \label{eq:a0}
\mathbf{A}_o = 
\left(
\begin{array}{c c c}
1     &  1     & 1     \\
1.1353 &  2.8133 & 0.5485 \\
1.2241 & 10.8140 & 0.2464 \\
1.2570 & 32.8359 & 0.1260 
\end{array}
\right) .
\end{equation}

\begin{figure} 
\epsfxsize=120mm
\begin{center}
\epsffile{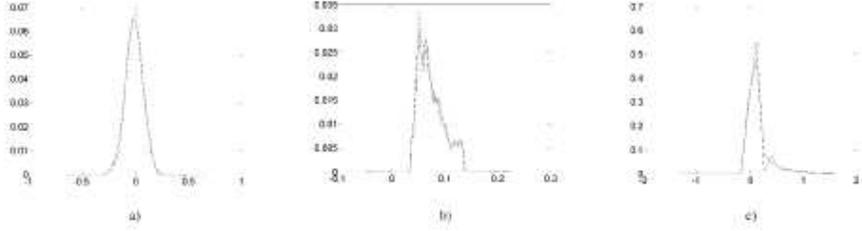}
\end{center}
\caption{Real (dotted) and estimated (solid) source density functions 
for a) CMB; b) synchrotron; c) dust.}
\label{fig:fig4}
\end{figure}

In figure~\ref{fig:fig2}, we show the data maps 
for stationary noise.  Also, note that the case 
examined does not fit the ICA assumptions.  For example, the 
normalized 
source covariance matrix at zero shift is:
\begin{equation}  \label{eq:cs}
\mathbf{C}_s(0,0) = 
\left(
\begin{array}{c c c }
	  1.0000     & 0.1961   &    0.0985  \\
	  0.1961   & 1.0000  &  0.6495	\\			
	0.0985 &   0.6495  &  1.0000
\end{array}
\right) 
\end{equation}
where a significant correlation, of the order of $65\%$, can be
observed between the dust and synchrotron maps.
 
For the data described above, we ran our learning algorithm for 500 
different noise realisations; for each run, 10000 iterations of the 
minimisation procedure described in the previous section were 
performed.  The unknown parameters were the spectral indices $n_s$ and 
$m$, and all the elements of matrices ${\bf C_{s}е}(\tau, \psi)$.  The 
cost defined in (\ref{eq:estimation_acs}), as a function 
of the iteration number in a particular run, is shown in 
figure~\ref{fig:fig3}.  The typical elapsed times per run were a few 
minutes on a 2 GHz CPU computer, with a Matlab interpreted code.  In 
the case described here, we estimated $n_s=2.8985$ and $m=1.7957$, corresponding 
to the mixing matrix
\begin{equation}  \label{eq:a_estimated}
\mathbf{A} = 
\left(
\begin{array}{c c c }
	1      &       1  &          1  \\
	1.1353 &   2.8118  &  0.5494	\\
	1.2241 &  10.8009  &  0.2473 \\
	1.2570  & 32.7775  &  0.1267 \\
\end{array}
\right) .
\end{equation}
As a quality index for our estimation, we adopted 
the matrix 
$\mathbf{Q}
\mathnormal{=}
(\mathbf{A}^{\mathrm{T}}\mathbf{C}_n^{\mathnormal{-1}}
\mathbf{A})^{\mathnormal{-1}}$ $(\mathbf{A}^{\mathrm{T}}\mathbf{C}_n^{\mathnormal{-1}}
\mathbf{A}_o)$,
which, in the ideal case, should be the 
$N \times N$ identity matrix $\mathbf{I}$. 
In the present case, we have:	
\begin{equation}  \label{eq:qualityQ}
\mathbf{Q}
=
\left(
\begin{array}{c c c }
	  1.0000  & -0.0074 &  -0.0013 \\
	 0.0000   & 1.0020  &  0.0000	\\
	  0.0000       &      0.0054  &  1.0013
\end{array}
\right) .
\end{equation}
The Frobenius norm of matrix 
$\mathbf{Q} \mathnormal{-} \mathbf{I}$
should be zero in the case of perfect model learning. 
In this case, it is $0.0096$. 

These results have been found by considering 25 uniformly distributed 
shift pairs, with $0 \leq \tau \leq 20$ and $0 \leq \psi \leq 20$.  As 
a synthetic index for the quality of the reconstructed source 
covariance matrices, we adopted a matrix $\bf E$, where each element 
is the relative error in the same covariance element, averaged over 
all the pairs $(\tau,\psi)$:
\begin{equation}  \label{eq:coverror}
\mathbf{E}_{i,j}е = \frac{1}{N_{s}ее+1}е \sum_{\tau, 
\psi}^{е}\frac{|\hat{\mathbf{C}е}_{\bf s}е_{i,j}ее(\tau,\psi) - 
\mathbf{C_{s}е}_{i,j}ее(\tau,\psi)|}{|\mathbf{C_{s}ее}_{i,j}ее(\tau,\psi)|}ее
\end{equation}
where $N_{s}е$ is the total number of shift pairs and 
$\hat{\mathbf{C}е}_{\bf s}е$ are the estimated source covariance 
matrices.  Of course, matrix (\ref{eq:coverror}) is only defined when 
all the denominators are nonzero.  A more accurate analysis of the 
results can be made from the element-by-element comparison 
of the estimated and the original matrices, but we do not report these 
results here.  For the case shown above, we have:
\begin{equation}  \label{eq:qualityE}
\mathbf{E}
=
\left(
\begin{array}{c c c }
	  0.0274  &  0.0392 &   0.0496 \\
	 0.0472  &  0.0170  &  0.0120	\\
	 0.0917  &  0.0125  &  0.0050
\end{array}
\right) .
\end{equation}

The reconstructed probability density functions of the source processes, 
estimated from equations 
(\ref{eq:y_element}) and (\ref{eq:convolution}), are shown in  
figure~\ref{fig:fig4}.

\begin{figure} 
\epsfxsize=120mm
\begin{center}
\epsffile{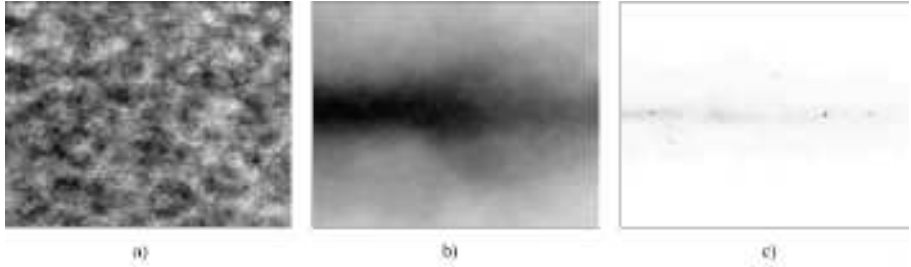}
\end{center}
\caption{
Wiener-filtered estimated maps: a) CMB; b) synchrotron; c) dust.}
\label{fig:fig5}
\end{figure}

\begin{figure} 
\epsfxsize=120mm
\begin{center}
\epsffile{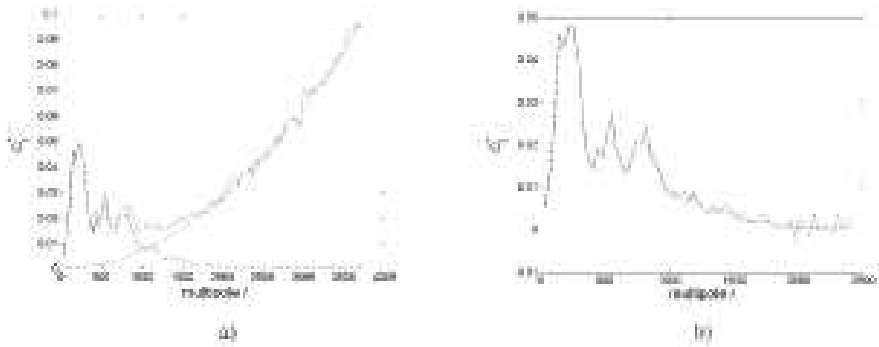}
\end{center}
\caption{a) Real (dotted) and estimated (solid) CMB power spectra.
The dashed line represents the theoretical power spectrum
of the noise component $n_{t_1}$  in (\ref{eq:y_element}),
evaluated from the noise covariance and the Moore-Penrose
pseudoinverse of the estimated mixing matrix. b) Real (dotted)
and estimated (solid) CMB power spectrum, corrected for theoretical
noise.}
\label{fig:fig6}
\end{figure}

We separated the sources by multiplying the 
data matrix by the Moore-Penrose generalised inverse, as in 
(\ref{eq:bmatrix2}), and then by applying a Wiener filter to the 
results thus obtained. As already said, this is not the best 
choice reconstruction algorithm at all, especially when 
the data are particularly noisy. However, the results 
we obtained are visually very good, as shown in figure~\ref{fig:fig5}. 
To evaluate more quantitatively the results of the whole 
learning-separation procedure, we compared the power spectrum 
of the CMB map with the one of the reconstructed map. 
This comparison is shown in Figure~\ref{fig:fig6}, where we also 
show the possibility of correcting the reconstructed 
spectrum for the known theoretical spectrum of the noise component 
$n_{t_1}$, obtained as in (\ref{eq:y_element}). 
As can be seen, the reconstructed spectrum is very 
similar to the original within a multipole $l=2000$.

\begin{figure} 
\epsfxsize=84mm
\begin{center}
\epsffile{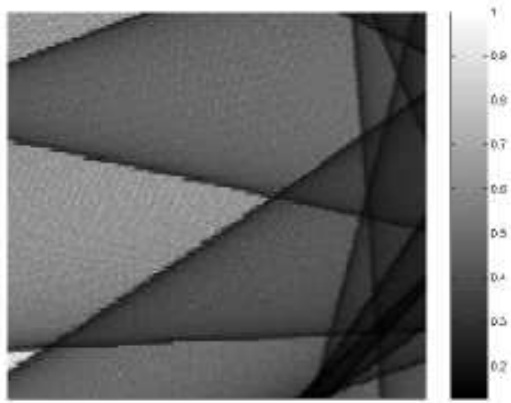}
\end{center}
\caption{Map of noise standard deviations used to generate 
nonstationary data}
\label{fig:fig7}
\end{figure}

\begin{figure} 
\epsfxsize=120mm
\epsffile{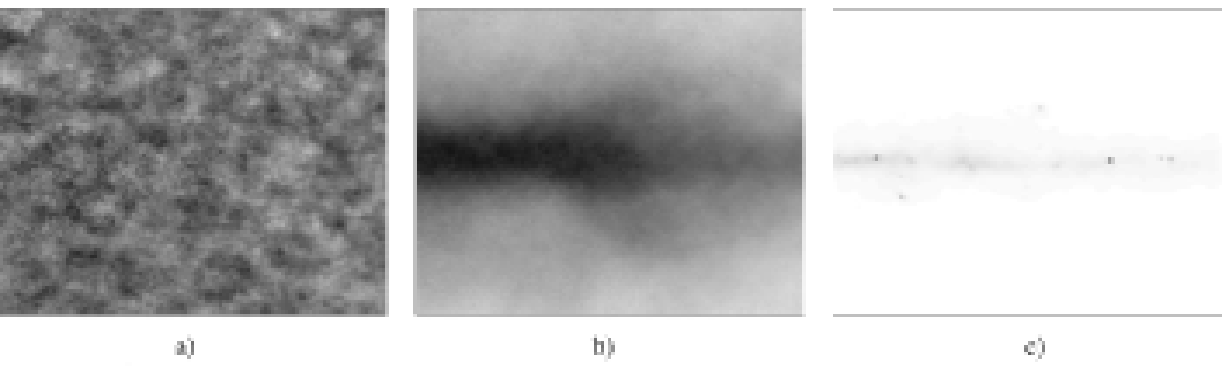}
\caption{Wiener-filtered estimated maps from nonstationary data: a) 
CMB; b) synchrotron; c) dust.}
\label{fig:fig8}
\end{figure}

Strictly speaking, our algorithm could not be applied to 
nonstationary processes. However,  let us assume that the original 
sources are stationary, and the noise is nonstationary but still 
spatially white and uncorrelated. This means that its covariance 
matrix, let us call it $\mathbf{R_n}(\xi,\eta)$, depends on the 
pixel. From our assumptions, these matrices are zero for any nonzero 
pair $(\tau,\psi)$.
We tried our method on nonstationary data, by assuming to know ${\bf 
R_{n}е}(\xi,\eta)$, and using a covariance matrix given by
\begin{equation}\label{eq:avgcov}
\mathbf{C_{n}е}(0,0) = 
\frac{1}{N_{p}ее}е\sum_{\xi,\eta}^{е}\mathbf{R_{n}е}(\xi,\eta)е
\end{equation}
The nonstationary data were obtained from a spatial template of noise 
standard deviations expected for typical Planck observations, shown in 
Figure \ref{fig:fig7}.  The actual standard deviations were adjusted 
so as to obtain the average signal-to-noise ratios desired for the 
different channels.  The separation results for a case where these 
SNRs were the same as in the above stationary case are 
shown in Figure \ref{fig:fig8}, where the degradation in the 
reconstruction is apparent in the regions where the noise is stronger.  
The results, in terms of recontructed power spectra, are perfectly 
comparable to the ones exemplified in Figure \ref{fig:fig6}.  The 
estimated spectral indices were $n_s=2.8885$ and $m=1.7881$, 
corresponding to the mixing matrix
\begin{equation}  \label{eq:a_estim_{var}е}
\mathbf{A} = 
\left(
\begin{array}{c c c }
	1      &       1  &          1  \\
	1.1353   & 2.8018  &  0.5509	\\
	1.2241   & 10.7128  &  0.2488 \\
	1.2570   & 32.3861  &  0.1279 \\
\end{array}
\right) .
\end{equation}
The average error on covariance matrices is in this case:
\begin{equation}  \label{eq:quality_{var}е}
\mathbf{E}
=
\left(
\begin{array}{c c c }
	  0.0158  &  0.1165  &  0.1930 \\
	 0.1163   & 0.0331   & 0.0254	\\
	 0.2440   & 0.0261   & 0.0144
\end{array}
\right) .
\end{equation}
The Frobenius norm of matrix $\mathbf{Q} \mathnormal{-} \mathbf{I}$ is 
now 0.0736, that is, slightly worse than for the above stationary case.

\section{Concluding remarks}
By exploiting the spatial structure of the sources, we developed an 
identification and separation algorithm that is able to exploit 
any available information on possible structure of the mixing matrix 
and the source covariance matrices. This can include the fully blind 
approach and the case exemplified here, where the mixing matrix is 
known to only depend on two parameters. The identification task is performed by a 
simple optimization strategy, while the proper separation can be 
faced by different approaches. We experimented the simplest one, but 
we are also developing more accurate techniques, especially suited to 
treat nonstationary noise on the data.

Our method is suitable to work directly with all-sky maps, but it 
could be necessary to apply it to small patches, as is shown in the 
above experimental section, to cope with the expected variability of 
the spectral indices and the noise variances in different sky regions.

It has been observed that it does not make sense to try source 
separation in those regions where the foreground emissions are much 
smaller than CMB and well below the noise level.  In any 
case, the CMB angular power spectrum has always been estimated fairly 
well up to a multipole $l=2000$, irrespective of the galactic 
latitude.  The estimation of the source densities has also given 
good results.  Source separation by our method has been particularly 
interesting with data from low galactic latitudes, where the 
foreground variance is often higher than the one of the CMB signal.  
Note that many separation strategies, both blind and non-blind, have 
failed their goal in this region of the celestial sphere.  As an 
example, WMAP data analysis (see \cite{bennett03}) was often performed 
by using pixel intensity masks that exclude the brightest sky portion 
from being considered.  Another interesting feature of our method is 
that significant cross-correlations between pairs of foregrounds can 
be straighforwardly taken into account.  Recently, some methods for a 
completely blind separation of correlated sources have been proposed 
in the literature (see for example \cite{barros00}).  Their 
effectiveness in astrophysical map separation has not been proved yet.  
Moreover, they have a high computational complexity.

Recently \cite{delabrouille02}, a frequency domain implementation of the 
method in \cite{belouchrani97} has been proposed.  
This method allows to take antenna beam effects into account 
straightforwardly by including the effect of the antenna 
transfer functions in the model.  It is also very flexible in introducing prior 
information about the entries of the mixing matrix and the spatial 
power spectra of the components.  An open problem is the extension of 
these methods to the case of correlated sources.  A possible extended 
method might be implemented in the space or in the frequency domain 
according to convenience.  Another problem that is still open with the 
expected Planck data is the different resolution of the data maps in 
some of the measurement channels.  The identification part of our 
method can work with maps whose resolution has been degraded in order 
to be the same in all the channels.  The result should be an estimate 
of the mixing matrix, which can be used in any non-blind separation 
approach with channel-dependent resolution, such as maximum entropy 
\cite{hobson98}.  
However, the possible asymmetry of the telescope beam patterns should 
be taken into account in verifying this possibility.

\section*{Acknowledgements}

This work has been partially supported by the Italian Space Agency, 
under contract ASI/CNR 1R/073/01. 
DH is supported by the
European Community's Human Potential Programme
under contract HPRN-CT-2000-00124 CMBNET. The authors adopted the simulated 
sky templates provided by the
Planck Technical Working Group 2.1 (diffuse component separation).
In particular, the authors are grateful to Martin Reinecke (MPA), Vlad
Stolyarov (Cambridge), Andrea Moneti, Simon Prunet and Fran\c{c}ois Bouchet
(IAP) for setting up and distributing the database. 
Extensive use of the HEALPix scheme 
(Hierarchical, Equal Area and iso-latitude pixelisation of the
sphere,\\ http://www.eso.org/science/healpix), by Krysztof M. 
G\'orski et al, has been made throughout this work.

\end{document}